\documentclass[prb,aps,showpacs,twocolumn]{revtex4}
\usepackage{times}
\usepackage{graphicx}
\usepackage{float}
\usepackage{latexsym,amsmath,amssymb,bm,euscript}
\usepackage{color}
\usepackage{subfigure}
\usepackage{epstopdf}
\usepackage[colorlinks=true,linkcolor=blue,citecolor=blue]{hyperref}
\usepackage{hyperref}

\begin{document}

\title{Many-body localization and mobility edge  in a disordered  Heisenberg spin ladder} 
\author{ Elliott Baygan, S. P. Lim and D.N. Sheng} 
\affiliation {Department of Physics and Astronomy, California State University, Northridge, California 91330, USA}

\begin{abstract} 
We examine the interplay of  interaction and disorder 
for a Heisenberg spin ladder system with random fields. 
We identify  many-body localized  states based on the entanglement entropy scaling, 
where  delocalized and localized states have volume and area laws, respectively.
We first establish the  quantum phase transition at a critical random field  strength $h_c \sim 8.5\pm 0.5$,
 where all  energy eigenstates are localized beyond that  value.
Interestingly, the entanglement entropy and  fluctuation of the bipartite magnetization
show distinct probability distributions which 
 characterize different quantum phases.  
Furthermore, we show that for  weaker $h$,  energy eigenstates with higher energy density are delocalized
while states at lower energy density are localized. This  defines a  mobility  edge  and a mobility gap 
 separating these two phases.  
By following the evolution of  low energy eigenstates, we observe that the  mobility gap  grows with 
increasing  the random field strength, which  drives the system to the phase of the full many-body localization
with increasing disorder strength.

\end{abstract}

\maketitle

\section{Introduction}

 Anderson localization  theory\cite{anderson1958} predicts that  noninteracting electrons are generally  
localized in one and two-dimensional (1D and 2D) disordered systems  without either a magnetic field 
or spin-orbit coupling due to  destructive quantum interference.
It is generally believed that low energy states remain  localized for weakly interacting systems
~\cite{basko2006, fleishman1980,altshuler1997,jacquod1997,georgeot1998,gornyi2005} with  characteristic features different from 
 noninteracting systems.  Recently, there is  renewed interest to examine
the Anderson localization for  interacting systems,
where the phenomenon of many-body localization (MBL)\cite{nandkishore2015,altman2015}
 has attracted  intense  studies.
Many remarkable properties of an MBL phase has been established\cite{
nandkishore2015,altman2015, nandkishore2014,oganesyan2007,pal2010,znidaric2008,
rigol2008, serbyn2014,kwasigroch2014,yao2014,vasseur2015,
huse2014, serbyn2013,ros2015,chandran2014,grover2014,
canovi2011,cuevas2012,bauer2013,kjall2014,luitz2015,luca2013,iyer2013,pekker_hilbert2014,johri2014,bardarson2012,andraschko2014,laumann2014,hickey2014,nanduri2014,barlev2014,imbrie2014,huang2015,you2015,serbyn2015,singh2015,barlev2015,deng2015,chen2015}
 based on  combined theoretical  and numerical studies.  For disordered interacting 
 systems, the random disorder can drive  a dynamic quantum phase transition\cite{nandkishore2015,pal2010,potter2015trans} from 
a delocalized state to an MBL phase,  where   energy eigenstates at  finite energy density become   localized.
From the quantum information perspective, energy  eigenstates in an MBL phase have suppressed
 entanglement entropy satisfying an area law\cite{nandkishore2015, bauer2013,kjall2014,  grover2014}
scaling with  the subsystem’s boundary area in contrast to the volume law scaling
expected for an ergodic   delocalized state.  As a consequence, 
the MBL phase is non-ergodic and can not thermalize\cite
{deutsch1991,srednicki1994,rigol2008}, which also challenges the fundamental ``eigenstate  thermalization hypothesis'' (ETH)  for quantum statistical  physics\cite{hosur2015}. The MBL state may exhibit  quantum order or    topological 
order\cite{huse2013,bahri2013,chandran2014,bauer2013,vosk2014,pekker_hilbert2014, potter2015,yao2015}
  at finite temperature as excitations at finite  energy  density are localized.
A phenomenological study~\cite{huse2014} further establishes that  the MBL
phase behaves like  integrable systems, respecting   extensive numbers  of local conservation laws 
~\cite{serbyn2013,ros2015,chandran2015}.
The quantum phase transition from an MBL phase to a delocalized ergodic phase 
may be continuous  characterized by a jump of  the entanglement entropy in the  thermodynamic limit\cite{grover2014},
where both entropy and its variance  grow with the system volume  at the critical point\cite{kjall2014,huse2014}.
Interestingly,  it is  conjectured that an MBL phase can also have a continuous localization-delocalization transition
to a new state, where the delocalized phase is non-ergodic whose volume law entanglement entropy tends to zero
as the transition is approached\cite{grover2014}.
It may  be possible to have the MBL phase in multi-component systems without random disorder\cite{groverf2014,ponte2015}.
The   MBL phase may be detected experimentally in cold atom systems 
~\cite{serbyn2014,kwasigroch2014,yao2014,vasseur2015}.

So far,  much of the  quantitative understanding  of  MBL systems are based on numerical 
 exact diagonalization (ED) studies\cite{oganesyan2007,pal2010,znidaric2008,canovi2011,cuevas2012,bauer2013,serbyn2014,kjall2014,luca2013,iyer2013,pekker2014,johri2014,bardarson2012,andraschko2014,laumann2014,vasseur2015,hickey2014,nanduri2014,barlev2014} 
 of     spin and electron systems, where  the dynamic  quantum phase transition between a 
delocalization phase  to an MBL  phase has been demonstrated   for 
different 1D model systems with spin (or particles) numbers in the range of $N=10-22$\cite{luitz2015, kjall2014}.  
 There are also some recent developments\cite{pekker2014_mps,friesdorf2015,pollmann2015,chandran2015_construct,khemani2015,yu2015}
 using tensor network and density matrix renormalization group approaches to study such systems.
One of the conceptually important and  unsettled issue is if the mobility edge exists for microscopic system to
separate the low energy localized state from the higher energy extended states.
On the one hand, these ED   studies\cite{luitz2015, kjall2014} have demonstrated the energy density dependence of the critical random field,  consistent with
the existence of the mobility edge. 
In particular,  Luitz et al.\cite{luitz2015} studied the 1D Heisenberg chain in a random field  using the  shift-inverted 
spectral transformation  method dealing with up to $22$ spins, where the finite-size scaling has been demonstrated with convincing
accuracy supporting the existence of the mobility edge.  However,  the recent numerical linked cluster  expansion 
study\cite{devakul2015} for a thermodynamic system  finds that  a  higher disorder strength 
(as the lower bound) is required to enter the MBL phase
  than that obtained by ED studies.  The reason for such a discrepancy remains not understood.
On the theoretical side,  it is not clear\cite{vosk_theory2014,roeck2015, laumann2014,shem2015, huang2015}   
 if some  spatial  region with higher energy density may play an important role with more extensive wavefunctions, 
 which may melt the lower energy eigenstates in the  system into delocalized states with  increasing
the  system size.
Some insight on this issue may  come from the earlier study of interacting many-body systems with random
disorder\cite{sheng2003,wan2005}, which exhibit
the fractionalized quantum Hall effect. In such a system,  we have demonstrated that 
 low energy states  below a mobility edge have  topological order protected by a mobility gap which separates the
low energy localized  insulating states  from the metallic states above the mobility gap.
These earlier studies suggest that it is possible  to follow the evolution of the low energy eigenstates in  disordered
interacting systems to detect if the mobility gap generally exists for MBL systems.

In this paper, we numerically examine the interplay of interaction and random disorder field for two-leg Heisenberg ladder systems,
which  stands between 1D and 2D systems\cite{barlev2015}, with the latter being  much harder to be  systematically studied based
on ED method.
We identify  MBL  states based on the bipartite entanglement entropy scaling,
and the spectral statistics of many-body energy levels.
We first establish the quantum phase transition at a critical random field  strength $h_c \sim 8.5\pm 0.5$,
 where all  energy eigenstates are localized beyond that  value.
Interestingly, the entanglement entropy shows distinct probability distribution in different quantum phases,
while transition is associated with the divergent variance for the entropy at the thermodynamic limit\cite{kjall2014, luitz2015}.
Despite the  small system sizes we can access with the number of spins $N=12-20$, our entropy distribution matches
to the theoretic prediction\cite{vosk_theory2014} in both delocalized  Griffiths phase and  MBL phase indicating 
we were able to access universal characteristics of these different quantum phases.
Furthermore, we show that at weaker $h$,  energy eigenstates with higher energy density are delocalized
while states at lower density are localized, which defines a  mobility  edge
 separating these two dynamically distinct quantum  phases   in agreement with earlier results for 1D spin chain systems\cite{kjall2014,luitz2015}.
The associated  mobility gap (the excitation gap of the lowest energy extended state to groundstate)
may be relevant for   experimental measurement  as it plays the role of the activation gap\cite{wan2005, sheng2007} for  
 quantum transport.
By following the evolution of  low energy eigenstates, we observe that the mobility edge moves to higher energy density  with the
increase of the random field strength, which eventually drives the system to the phase  with  full MBL where
all energy eigenstates are localized.
The remaining of the paper is organized as following:
In Sec. II, we first introduce the two-leg ladder spin model with random fields and briefly discuss the method
we use to study the system. 
We present the evidence of the MBL phase  determined by the entanglement entropy and the energy level statistics studies.
In Sec. III, we  study the characteristic features of the quantum phases and the transition between
the delocalized phase and the MBL phase.  We  also present the evidence for the mobility edge separating the low energy localized
phase from the higher energy extended states.
 Finally, in Sec. IV,  we summarize  our main results and discuss open questions.

\section{Spin model and transition to an MBL phase}

We study the Heisenberg two leg ladder spin system  on the square   lattice
with  the following Hamiltonian:
\begin{equation}
 H = J \sum_{\langle i,j \rangle} \vec{S}_i \cdot \vec{S}_j 
% + J'_1 \sum_{\langle i,j \rangle_{\rm zigzag}} \vec{S}_i \cdot \vec{S}_j 
  -  \sum_{i} h_iS_i^z,\nonumber
% H=\sum_{i\in [1,L]} {\bf S}_i \cdot {\bf  S}_{i+1} -h_iS_i^z, \label{eq:H} \ee
\end{equation}
where the summation $\langle i,j \rangle$ 
runs over all  distinct nearest neighbor bonds with antiferromagnetic coupling  $J$, which is set  as the units
of the energy $J=1$.
The $h_i$ is a random magnetic field coupling, which  distributes  uniformly between 
window $(-h,h)$ with $h$ as the strength of  random fields.
The number of sites  of the ladder system can be written as
$N=N_xN_y$ with $N_y=2$ and $N_x$ is the number of sites along each spin chain.

\begin{figure}[b]
 \includegraphics[width=2.4in]{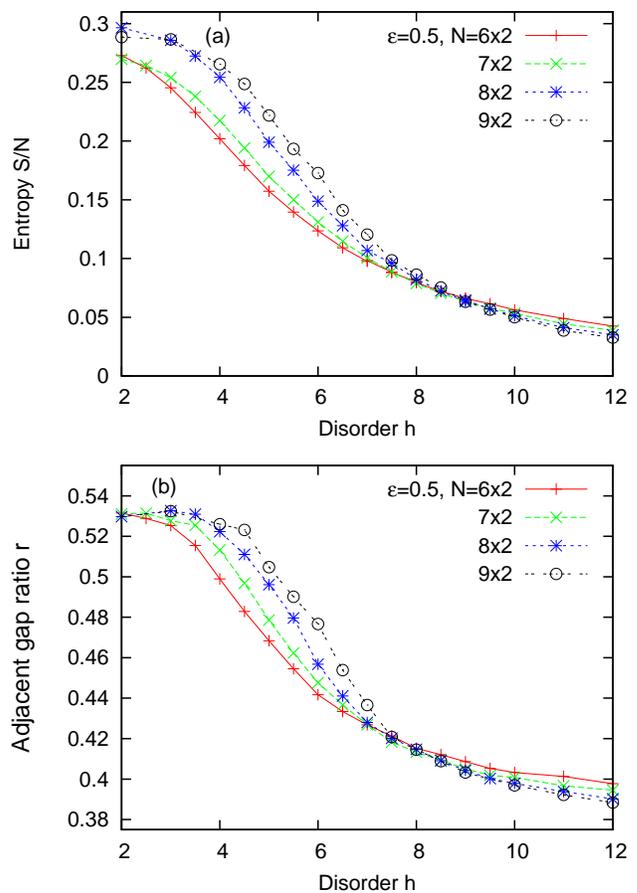}
\hspace{0.0in}
\vspace{1.0in}
\caption{(Color online) (a) The ratio of  entanglement entropy over the number of system sites  $S/N$
for  different systems  from
$N=6\times 2$ to $9\times 2$  at the energy density $\varepsilon =0.5$ as
a function of random disorder  strength $h$.
Curves for different $N$  approximately cross each other around  a critical random field strength  $h_c=8.5 \pm 0.5$.
(b) The adjacent gap ratio  $r$ 
for states with energy density $\varepsilon=0.5$ as a function of disorder strength $h$.  Here we see that on small
$h$ side, $r$ approaches  the Gaussian  orthogonal ensemble  value (0.5307) representing delocalized states, while at larger 
$h$ side,  $r$ reaches the Poisson value $(2ln2-1\simeq 0.3863)$ for larger systems representing  localized states.  
All curves cross around the  critical value $h_c = 8.0 \pm 0.7$.  
 }
\label{fig1}
\end{figure}

We  perform Lanczos ED calculations     to obtain  energy eigenstates around a fixed value
$E$ determined by the target  energy density $\varepsilon$ for  systems with the number of sites
$N=12-20$  in the total $S_z=0$ sector. 
Specifically, for each disorder configuration, we first calculate the 
ground state energy  $E_0$ and the maximum energy $E_\text{max}$, which are  used to define the target 
energy density $\varepsilon = (E-E_0)/(E_{max} -E_0)$.
We perform more than $1000$ disorder configuration average for most  systems we studied.
Physical quantities\cite{luitz2015}  including the bipartite entanglement entropy,  energy level statistics and 
bipartite  fluctuation of the subsystem magnetization are obtained and averaged over  
  disorder configurations and sometimes also averaged over  30 energy eigenstates with energies closest to the given
energy density $\varepsilon$ as detailed below.

\begin{figure}[b]
 \includegraphics[width=2.3in]{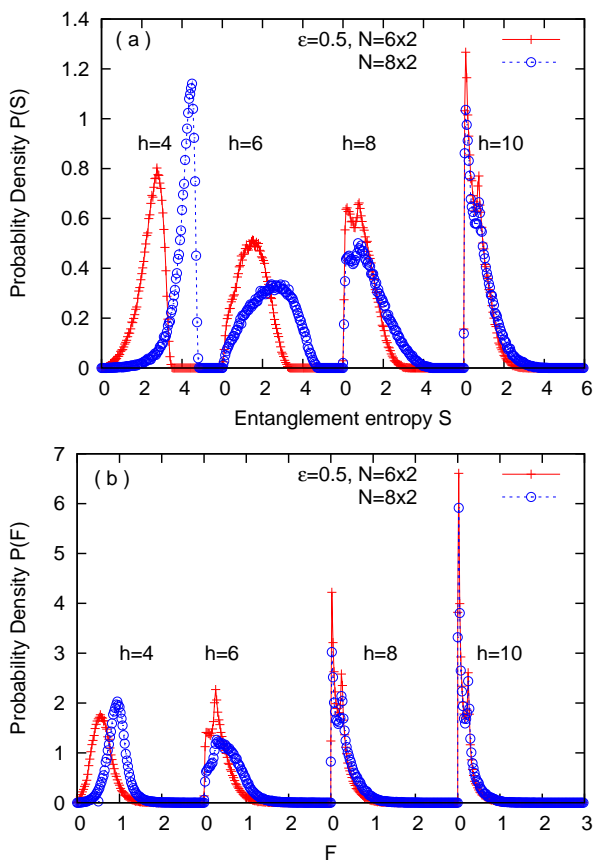}
\hspace{-0.0in}
\vspace{1.0in}
\caption{(Color online)  (a) The probability distributions  of  the bipartition entanglement  entropy $P(S)$ 
for spin system at energy density $\varepsilon=0.5$ with
different disorder strengths $h=4$, $6$, $8$ and $10$ for system sizes  $N=6\times 2$ and  $8\times 2$.
 These results illustrate that for stronger disorder case,  the distribution always has a long tail 
into higher $S$ values, while for smaller $h$ the long tail is on  small $S$ side.
(b)  The probability distributions $P(F)$  of the variance $F=<(S_{A}^z)^2>-<S_{A}^z>^2$ (in units
of the square of the  Planck constant $\hbar^2$)
of the magnetization of the half system A  for spin system at energy density $\varepsilon=0.5$ with
different disorder strengths $h=4$, $6$, $8$ and $10$ for
  $N=6\times 2$ and (b) $8\times 2$.
 }

\label{fig2prob1}
\end{figure}

The bipartite entanglement entropy has been extensively used as an effective tool  to
characterize  quantum phases for such an  interacting system\cite{nandkishore2015, luitz2015, kjall2014}
 We compute the Von Neumann entanglement entropy of the ladder system  from all eigenvalues of the reduced density matrix
$\rho_A$ as
 $S=- {\rm Tr} \rho_A \ln \rho_A$,
 by partitioning the system in  the middle
along the vertical direction (the lengths for  two subsystems A and B are the integer-parts of 
  $N_x/2$ and $(N_x+1)/2$, respectively). 
For an interacting  system with weak disorder,  
the entanglement entropies of   higher energy eigenstates are  expected to follow the
 volume law and these states are ergodic satisfying  the ETH\cite{nandkishore2015,altman2015}.  
This is in contrast to the behavior of the  ground state,  where
 the entanglement entropy follows the area law (with possibly up to the logarithmic correction
depending on if there are gapless excitations)\cite{grover2014}. 
By varying the disorder strength  $h$,  one can detect the possible quantum phase transition from the  behavior of the
entanglement entropy\cite{kjall2014, luitz2015}.
As shown in  Fig. 1(a), we plot the  ratio of  entanglement entropy over the number of system sites  $S/N$
for  different systems  from
$N_x=6$ to $9$  at the energy density $\varepsilon =0.5$ as
a function of random field strength $h$.  
On the  smaller $h$ side, we see the ratio $S/N$ increases with  system sizes $N$ and  approaches a constant  indicating the volume law growth of  $S$.  
With varying $h$,  all data points approximately cross each other around  a critical value  $h_c\sim 8.5\pm 0.5$.
On the  larger $h$ side,  $S/N$ approaches zero 
indicating the low entanglement and non-ergodic behavior  where  energy eigenstates are localized.
The ladder systems we study here have stronger finite size effect  (from the even-odd effect of $N_x$)
than the 1D spin chain systems, which is the reason that not all curves cross at the same point in Fig. 1(a).  

\begin{figure}[b]
  \includegraphics[width=2.3in]{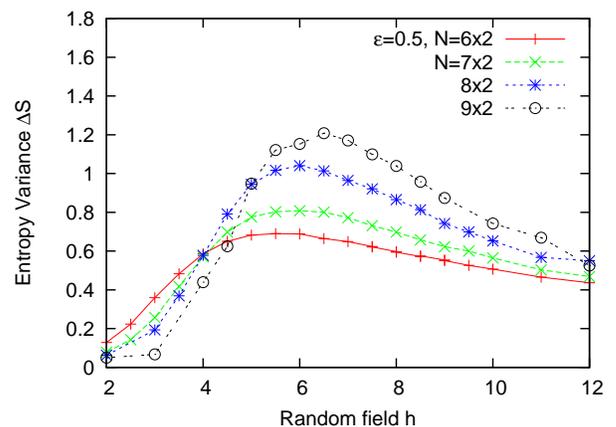}
\hspace{-0.0in}
\vspace{-1.1in}
\caption{(Color online) (a) The variance $(\Delta S)^2= <S^2> -<S>^2$ of the entanglement entropy  
at  energy density $\varepsilon=0.5$ for different $h$.
$\Delta S$ reaches the peak value at $h_p$  smaller than the identified $h_c$ for quantum phase transition.
Clearly,  $h_p$   shifts to higher $h$ with the increase of $N$.
 }
\label{fig3_val}
\end{figure}

\begin{figure}[b]
 \includegraphics[width=2.in]{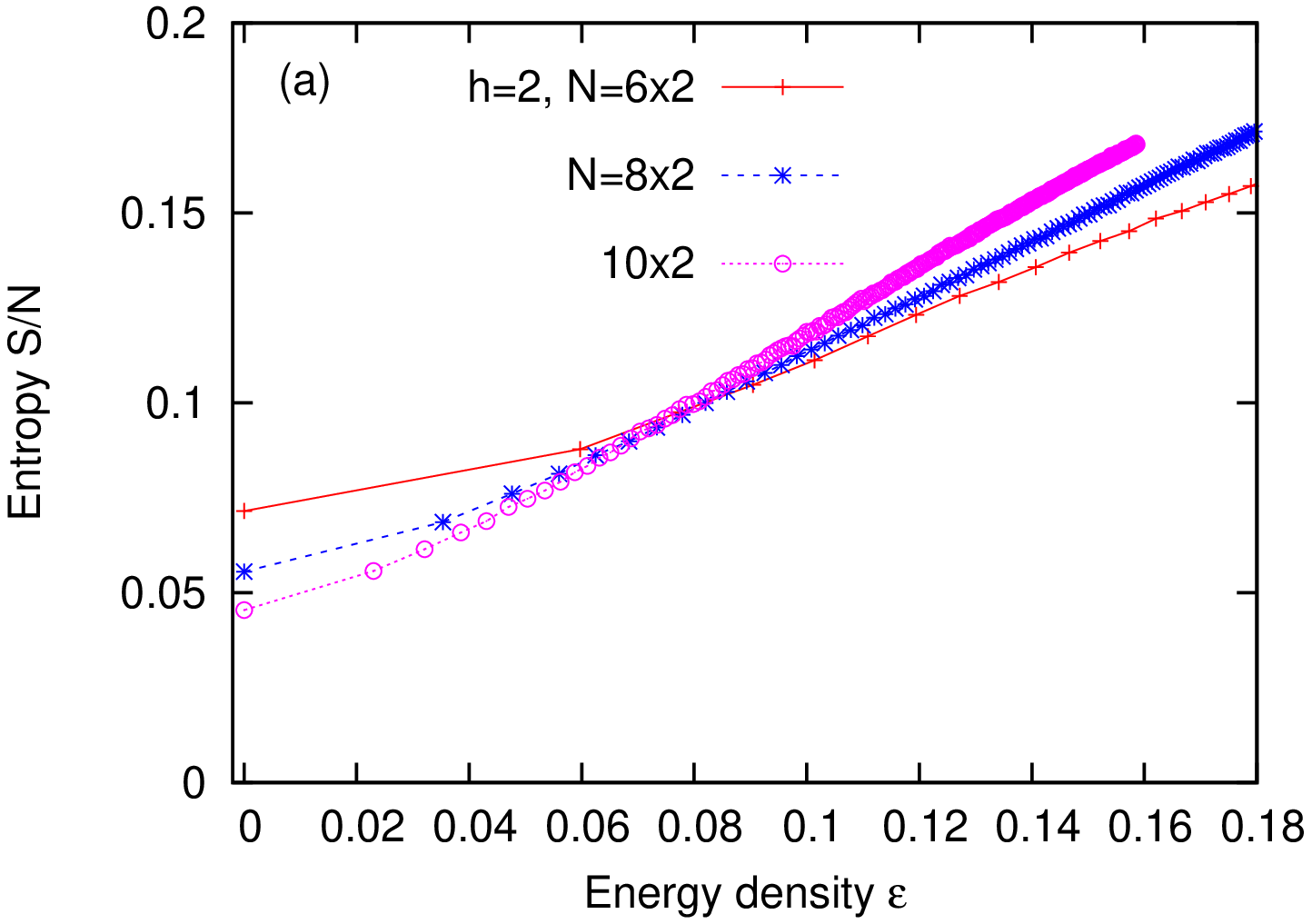}
 \includegraphics[width=2.in]{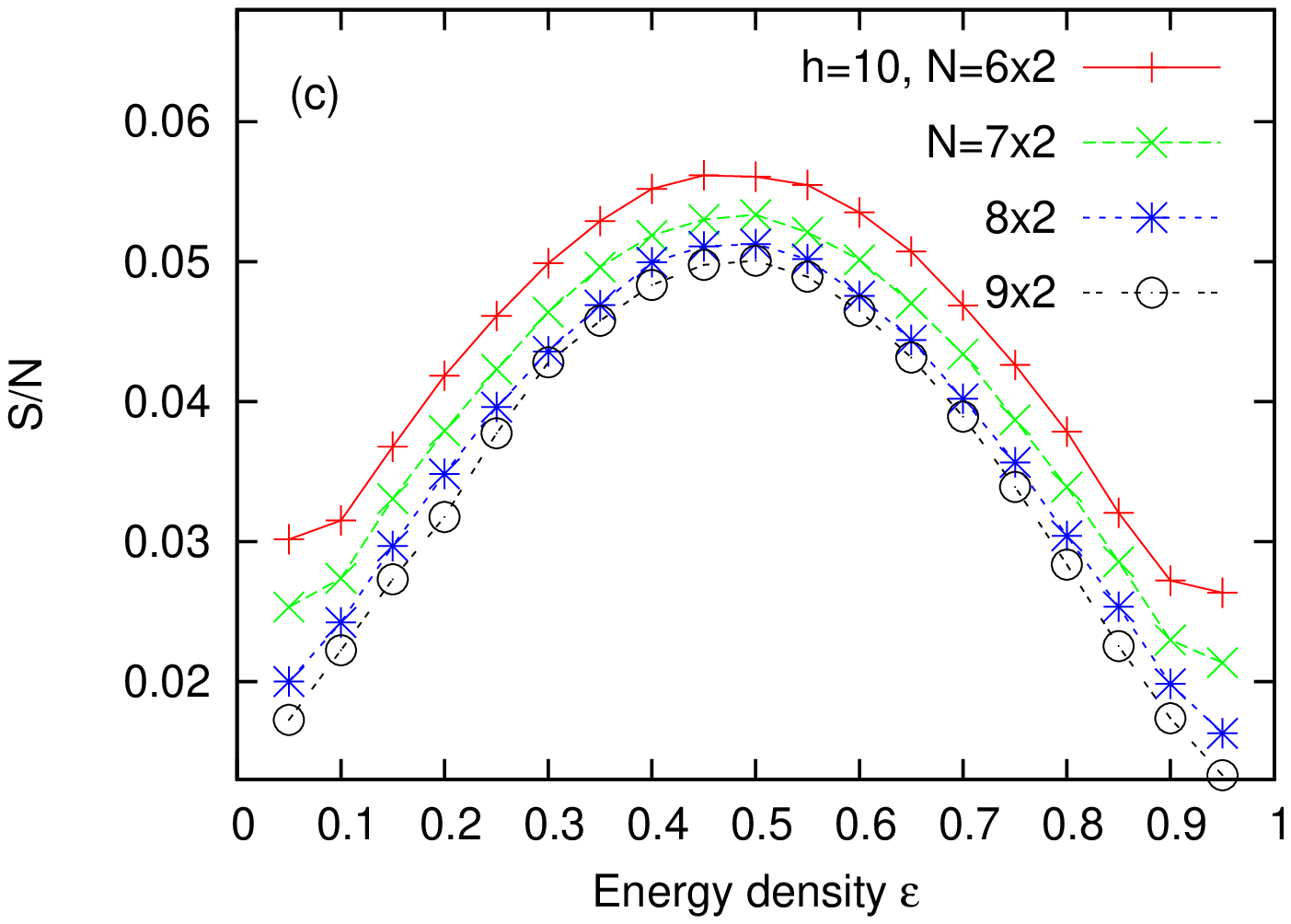}
\hspace{-0.0in}
\vspace{0.3in}
\caption{(Color online)  The entropy of each energy eigenstate $S_i$  for  low energy eigenstates
 averaged over disorder configurations is shown as a function of its average energy density.   
   $S_i/N$ curves for different system sizes  $N=6\times 2$,  $8\times 2$ and $10\times 2$ cross at a critical
energy density $\varepsilon_c$, which separates the higher energy delocalized states from the
lower energy localized states. 
 }
\label{thouless1}
\end{figure}

We further use the  level statistics analysis from the random matrix theory\cite{atas2013,oganesyan2007}
  to probe the localization-delocalization characteristics  
of  energy  eigenstates. 
In the delocalized  regime, the level-spacing  distribution 
is described by the   Gaussian orthogonal ensemble (GOE) statistics, which
represents extended levels with level-repulsion  between them because
of the overlap of  energy eigenstates in real space.
In the localized regime, the level-spacing distribution  is 
determined by Poisson statistics as  wave-functions
close in energy are exponentially localized with no 
level repulsion between them\cite{mehta1991}.
In the energy spectrum analysis\cite{luitz2015}, 
we define the energy gap $\delta_n=E_n-E_{n-1}$ as the energy difference between the $n$-th and $(n-1)$-th eigenstates,
then the adjacent gap ratio can be defined as 
$r_n=min(\delta_n, \delta_{n+1})/max(\delta_n, \delta_{n+1})$.
We  average the gap ratio $r=<r_n>$  over  states near the spectrum center
at $\varepsilon=0.5$ for 30 eigenstates and 1000 random disorder configurations
for each given disorder  strength $h$.   As shown in Fig. 1(b),   we see that at the small
$h$ side, $r$ approaches  the Gaussian  orthogonal ensemble  value (0.5307) representing delocalized states, while at stronger
$h$ side,  $r$ reaches the Poisson value $(2ln2-1\simeq 0.3863)$ for larger systems representing the level statistics of   localized states.  
All curves cross around the  critical value $h_c \sim 8.0-8.5$.  Due to the stronger finite size effect
coming from the even-odd effect  for  $N_x$, we are not attempting to do a finite-size
scaling. Instead, we will focus on the  universal behavior of the different phases to explore  the  nature
of the quantum phases and the transition involved here.

\section{Nature of  quantum phases and phase transition}

\subsection{Probability distributions  of  entanglement entropy and   variance of   bipartite magnetization}

Now we turn to the study of the  probability distribution  of the entropy $P(S)$ 
for spin system at energy density $\varepsilon=0.5$ for
different disorder strengths $h=4$, $6$, $8$ and $10$ crossing two different phases obtained
for  ensembles with 30 energy eigenstates and 1000 disorder configurations for each $h$.
As shown in the Fig. 2(a), 
 on the  small $h$ side, we see that the peak position of the  distribution $P(S)$ (which
reflects the average of $S$)  moves to the larger $S$ value  with  increasing
system size $N$, indicating a consistency with the volume law for the entropy.
Close to the transition point for $h=6$ we find that the distribution $P(S)$ becomes
much broadened while the peak position  moves to the higher $S$ with the increase of $N$, but the peak height 
reduces at the same time.   As we move to the higher $h$ side,  we see that the distribution again
becomes sharper, with two peaks showing for each $P(S)$ curve which may be related to  the non-ergodic character
of the localized phase.
 Furthermore, we also see that  for the stronger disorder case,  the distribution always has a long tail 
into higher $S$ values, while for smaller $h$ the long tail is at the smaller $S$ side.
To compare with recent theoretical description of the MBL of 1D system\cite{vosk_theory2014},  we find that the
entropy distribution is very similar to the ones obtained based on their real space renormalization group simulations.
Specifically, at $h=4$ our distribution has a long tail in  small $S$ region, which matches to the one for the
Griffiths phase just before the transition to the MBL phase. At  $h=10$, we find the $P(S)$ is peaked
at $S=0$ and shows an exponential decay tail on the larger $S$ side.

We compare the entanglement entropy behavior with the  bipartite fluctuations $F$ of the subsystem
magnetization $S^z_A$~\cite{luitz2015,song2012}, which is defined as $F=\langle \left( S^z_A \right)^2 \rangle
- \langle S^z_A \rangle^2$ as shown  in Fig. 2(b).
The  distribution $P(F)$ exhibits very similar behavior  as $P(S)$ for $h$ closer to  the quantum phase
transition and in the MBL phase. 
The similar two-peaks structure is also clear  for $P(F)$ on the  larger $h$ side in the MBL phase.
The only difference worth mentioning is that with  weaker disorder  $h=4$,  the $P(F)$  demonstrates
the normal Gaussian distribution,  which is sharp and near symmetric about the peak.

The variance of the entanglement entropy has been shown to be  
an excellent quantity\cite{kjall2014, vosk_theory2014}
 for identifying the quantum phase transition from  1D spin chain studies. Here we 
show the variance $\Delta S$ of the entanglement entropy  
averaged over  $30$ different energy eigenstates  around the energy density $\varepsilon=0.5$
and   1000 disorder configurations. 
In agreement with these observations for 1D systems\cite{kjall2014, luitz2015}, we find that the
$\Delta S$  is small in both small $h$ and large $h$ sides,  and demonstrates a peak for the intermediate
$h$ as shown in Fig. 3.   
We  observe that the peak value of $\Delta S$ increases with $N$, which may diverge 
at the transition point.
The   position of the peak $h_p$ is smaller than the previously identified
$h_c$ and it shows a trend of approaching $h_c$ with the increase of $N$.
These results are consistent with the phenomenological theory\cite{vosk_theory2014} established  based on the real
space renormalization group studies, which indicate that we are observing  intrinsic properties
of the MBL phase and the related quantum phase transition  for  system sizes we study.

\subsection {Mobility edge and  mobility gap}

We have shown that the disorder can drive a quantum phase transition, where all states near the center of energy
spectrum are localized.  In fact,   all other states with different energy density are also localized,
 thus we enter the full MBL phase tuned by  $h$ (see Fig. 4(c) as an example).
To address the issue if the mobility edge naturally exists in such a system separating  low energy
states with the area-law entanglement entropy from  higher energy states with  volume-law behavior,
we follow  lower energy eigenstate by obtaining hundreds of  these eigenstates in ED.
 We follow the entropy of each energy eigenstate $S_i$  in the lower energy
density regime and average that over 1000 disorder configurations.

As shown in Fig. 4(a-b) for $h=2$ and $4$, 
we identify that the disorder configuration averaged entropy $S_i$ for the $i-th$ energy eigenstate is  a smooth increasing function of eigen energy $E_i$ or its average energy
density $\varepsilon_i = <(E_i- E_0)/(E_{max}-E_0)>$.
The entropy per site   $S_i/N$ for different system sizes $N=6\times 2$,  $8\times 2$ and $10\times 2$ crosses around  a critical
energy density $\varepsilon_c$, which separates  higher energy states with volume-law entropy (delocalized states) from 
lower energy localized states with $S_i/N $ approaching zero violating the volume law. 
Here all the data we show have  even $N_x=6$, $8$ and $10$ with reduced finite size effect.  The crossing point  determines the mobility edge.
  With increasing $h$,  the  entropy of the low-lying eigenstate  is reduced
and the mobility edge is being pushed to the higher energy density from $\varepsilon_c\sim 0.075$ at $h=2$ to
$\varepsilon_c\sim 0.1$ at $h=4$.   
As shown in Fig. 4(c),  we further move to the stronger disorder case, at $h=10$, we see that $S/N$  at different energy density (here we averaged over both
disorder configurations and energy  eigenstates for each energy density $\varepsilon$) is always a decreasing function
with  increasing  $N$ demonstrating all states are localized.

\section{Summary and discussion}

We have identified the disorder driven dynamic quantum phase transition from ergodic delocalized phase
to an MBL non-ergodic phase for two-leg ladder Heisenberg spin systems  with random  field disorder. 
The characteristic distributions of the entanglement entropy for both  the delocalized Griffiths  phase  and the MBL
phase  agree with the theoretical description for the
MBL\cite{vosk_theory2014}.   
Furthermore, we show that for  weaker $h$,  energy eigenstates with higher energy density are delocalized,
while states at lower density are localized, which defines a  mobility  edge
 separating these two dynamically distinct quantum  states  in agreement with earlier results for 1D spin chain systems\cite{kjall2014,luitz2015}.
On the quantitative side, we  find that the Heisenberg ladder requires a much higher critical disorder strength compared to
the 1D spin chain model\cite{luitz2015} to enter the full MBL phase.  It is highly desired to further study multi-leg ladders and explore
 scaling behaviors to the 2D limit, which we  leave  for a future study.

{\bf Acknowledgments} - We thank Tarun Grover and David Huse for stimulating discussions.
This work is supported by US National Science Foundation  Grants 
PREM DMR-1205734 (EB), DMR-1408560, and Princeton MRSEC Grant  DMR-1420541  for travel support.

%%%%%%%%%%%%%%%%%%%%%%%%%%%%%%%%%%%% References
%\bibliographystyle{apsrev}
\bibliography{randomfield}

\begin{thebibliography}{75}
\expandafter\ifx\csname natexlab\endcsname\relax\def\natexlab#1{#1}\fi
\expandafter\ifx\csname bibnamefont\endcsname\relax
  \def\bibnamefont#1{#1}\fi
\expandafter\ifx\csname bibfnamefont\endcsname\relax
  \def\bibfnamefont#1{#1}\fi
\expandafter\ifx\csname citenamefont\endcsname\relax
  \def\citenamefont#1{#1}\fi
\expandafter\ifx\csname url\endcsname\relax
  \def\url#1{\texttt{#1}}\fi
\expandafter\ifx\csname urlprefix\endcsname\relax\def\urlprefix{URL }\fi
\providecommand{\bibinfo}[2]{#2}
\providecommand{\eprint}[2][]{\url{#2}}

\bibitem[{\citenamefont{{Anderson}}(1958)}]{anderson1958}
\bibinfo{author}{\bibfnamefont{P.~W.} \bibnamefont{{Anderson}}},
  \bibinfo{journal}{Phys. Rev.} \textbf{\bibinfo{volume}{109}},
  \bibinfo{pages}{1492} (\bibinfo{year}{1958}).

\bibitem[{\citenamefont{{Basko} et~al.}(2006)\citenamefont{{Basko}, {Aleiner},
  and {Altshuler}}}]{basko2006}
\bibinfo{author}{\bibfnamefont{D.~M.} \bibnamefont{{Basko}}},
  \bibinfo{author}{\bibfnamefont{I.~L.} \bibnamefont{{Aleiner}}},
  \bibnamefont{and} \bibinfo{author}{\bibfnamefont{B.~L.}
  \bibnamefont{{Altshuler}}}, \bibinfo{journal}{Annals of Physics}
  \textbf{\bibinfo{volume}{321}}, \bibinfo{pages}{1126} (\bibinfo{year}{2006}).

\bibitem[{\citenamefont{{Fleishman} and {Anderson}}(1980)}]{fleishman1980}
\bibinfo{author}{\bibfnamefont{L.}~\bibnamefont{{Fleishman}}} \bibnamefont{and}
  \bibinfo{author}{\bibfnamefont{P.~W.} \bibnamefont{{Anderson}}},
  \bibinfo{journal}{\prb} \textbf{\bibinfo{volume}{21}}, \bibinfo{pages}{2366}
  (\bibinfo{year}{1980}).

\bibitem[{\citenamefont{{Altshuler} et~al.}(1997)\citenamefont{{Altshuler},
  {Gefen}, {Kamenev}, and {Levitov}}}]{altshuler1997}
\bibinfo{author}{\bibfnamefont{B.~L.} \bibnamefont{{Altshuler}}},
  \bibinfo{author}{\bibfnamefont{Y.}~\bibnamefont{{Gefen}}},
  \bibinfo{author}{\bibfnamefont{A.}~\bibnamefont{{Kamenev}}},
  \bibnamefont{and} \bibinfo{author}{\bibfnamefont{L.~S.}
  \bibnamefont{{Levitov}}}, \bibinfo{journal}{Phys. Rev. Lett.}
  \textbf{\bibinfo{volume}{78}}, \bibinfo{pages}{2803} (\bibinfo{year}{1997}).

\bibitem[{\citenamefont{{Jacquod} and {Shepelyansky}}(1997)}]{jacquod1997}
\bibinfo{author}{\bibfnamefont{P.}~\bibnamefont{{Jacquod}}} \bibnamefont{and}
  \bibinfo{author}{\bibfnamefont{D.~L.} \bibnamefont{{Shepelyansky}}},
  \bibinfo{journal}{Phys. Rev. Lett.} \textbf{\bibinfo{volume}{79}},
  \bibinfo{pages}{1837} (\bibinfo{year}{1997}).

\bibitem[{\citenamefont{{Georgeot} and {Shepelyansky}}(1998)}]{georgeot1998}
\bibinfo{author}{\bibfnamefont{B.}~\bibnamefont{{Georgeot}}} \bibnamefont{and}
  \bibinfo{author}{\bibfnamefont{D.~L.} \bibnamefont{{Shepelyansky}}},
  \bibinfo{journal}{Phys. Rev. Lett.} \textbf{\bibinfo{volume}{81}},
  \bibinfo{pages}{5129} (\bibinfo{year}{1998}).

\bibitem[{\citenamefont{{Gornyi} et~al.}(2005)\citenamefont{{Gornyi}, {Mirlin},
  and {Polyakov}}}]{gornyi2005}
\bibinfo{author}{\bibfnamefont{I.~V.} \bibnamefont{{Gornyi}}},
  \bibinfo{author}{\bibfnamefont{A.~D.} \bibnamefont{{Mirlin}}},
  \bibnamefont{and} \bibinfo{author}{\bibfnamefont{D.~G.}
  \bibnamefont{{Polyakov}}}, \bibinfo{journal}{Phys. Rev. Lett.}
  \textbf{\bibinfo{volume}{95}}, \bibinfo{eid}{206603} (\bibinfo{year}{2005}).

\bibitem[{\citenamefont{{Nandkishore} and {Huse}}(2015)}]{nandkishore2015}
\bibinfo{author}{\bibfnamefont{R.}~\bibnamefont{{Nandkishore}}}
  \bibnamefont{and} \bibinfo{author}{\bibfnamefont{D.~A.}
  \bibnamefont{{Huse}}}, \bibinfo{journal}{Annu. Rev. Cond. Matt. Phys.}
  \textbf{\bibinfo{volume}{6}}, \bibinfo{pages}{15} (\bibinfo{year}{2015}).

\bibitem[{\citenamefont{{Altman} and {Vosk}}(2015)}]{altman2015}
\bibinfo{author}{\bibfnamefont{E.}~\bibnamefont{{Altman}}} \bibnamefont{and}
  \bibinfo{author}{\bibfnamefont{R.}~\bibnamefont{{Vosk}}},
  \bibinfo{journal}{Annu. Rev. Cond. Matt. Phys.} \textbf{\bibinfo{volume}{6}},
  \bibinfo{pages}{383} (\bibinfo{year}{2015}).

\bibitem[{\citenamefont{{Nandkishore} et~al.}(2014)\citenamefont{{Nandkishore},
  {Gopalakrishnan}, and {Huse}}}]{nandkishore2014}
\bibinfo{author}{\bibfnamefont{R.}~\bibnamefont{{Nandkishore}}},
  \bibinfo{author}{\bibfnamefont{S.}~\bibnamefont{{Gopalakrishnan}}},
  \bibnamefont{and} \bibinfo{author}{\bibfnamefont{D.~A.}
  \bibnamefont{{Huse}}}, \bibinfo{journal}{\prb} \textbf{\bibinfo{volume}{90}},
  \bibinfo{eid}{064203} (\bibinfo{year}{2014}).

\bibitem[{\citenamefont{{Rigol} et~al.}(2008)\citenamefont{{Rigol}, {Dunjko},
  and {Olshanii}}}]{rigol2008}
\bibinfo{author}{\bibfnamefont{M.}~\bibnamefont{{Rigol}}},
  \bibinfo{author}{\bibfnamefont{V.}~\bibnamefont{{Dunjko}}}, \bibnamefont{and}
  \bibinfo{author}{\bibfnamefont{M.}~\bibnamefont{{Olshanii}}},
  \bibinfo{journal}{\nat} \textbf{\bibinfo{volume}{452}}, \bibinfo{pages}{854}
  (\bibinfo{year}{2008}).

\bibitem[{\citenamefont{{Serbyn} et~al.}(2014)\citenamefont{{Serbyn}, {Knap},
  {Gopalakrishnan}, {Papi{\'c}}, {Yao}, {Laumann}, {Abanin}, {Lukin}, and
  {Demler}}}]{serbyn2014}
\bibinfo{author}{\bibfnamefont{M.}~\bibnamefont{{Serbyn}}},
  \bibinfo{author}{\bibfnamefont{M.}~\bibnamefont{{Knap}}},
  \bibinfo{author}{\bibfnamefont{S.}~\bibnamefont{{Gopalakrishnan}}},
  \bibinfo{author}{\bibfnamefont{Z.}~\bibnamefont{{Papi{\'c}}}},
  \bibinfo{author}{\bibfnamefont{N.~Y.} \bibnamefont{{Yao}}},
  \bibinfo{author}{\bibfnamefont{C.~R.} \bibnamefont{{Laumann}}},
  \bibinfo{author}{\bibfnamefont{D.~A.} \bibnamefont{{Abanin}}},
  \bibinfo{author}{\bibfnamefont{M.~D.} \bibnamefont{{Lukin}}},
  \bibnamefont{and} \bibinfo{author}{\bibfnamefont{E.~A.}
  \bibnamefont{{Demler}}}, \bibinfo{journal}{Phys. Rev. Lett.}
  \textbf{\bibinfo{volume}{113}}, \bibinfo{eid}{147204} (\bibinfo{year}{2014}).

\bibitem[{\citenamefont{{Kwasigroch} and {Cooper}}(2014)}]{kwasigroch2014}
\bibinfo{author}{\bibfnamefont{M.~P.} \bibnamefont{{Kwasigroch}}}
  \bibnamefont{and} \bibinfo{author}{\bibfnamefont{N.~R.}
  \bibnamefont{{Cooper}}}, \bibinfo{journal}{\pra}
  \textbf{\bibinfo{volume}{90}}, \bibinfo{eid}{021605} (\bibinfo{year}{2014}).

\bibitem[{\citenamefont{{Huse} et~al.}(2014)\citenamefont{{Huse},
  {Nandkishore}, and {Oganesyan}}}]{huse2014}
\bibinfo{author}{\bibfnamefont{D.~A.} \bibnamefont{{Huse}}},
  \bibinfo{author}{\bibfnamefont{R.}~\bibnamefont{{Nandkishore}}},
  \bibnamefont{and}
  \bibinfo{author}{\bibfnamefont{V.}~\bibnamefont{{Oganesyan}}},
  \bibinfo{journal}{\prb} \textbf{\bibinfo{volume}{90}}, \bibinfo{eid}{174202}
  (\bibinfo{year}{2014}).

\bibitem[{\citenamefont{{Serbyn} et~al.}(2013)\citenamefont{{Serbyn},
  {Papi{\'c}}, and {Abanin}}}]{serbyn2013}
\bibinfo{author}{\bibfnamefont{M.}~\bibnamefont{{Serbyn}}},
  \bibinfo{author}{\bibfnamefont{Z.}~\bibnamefont{{Papi{\'c}}}},
  \bibnamefont{and} \bibinfo{author}{\bibfnamefont{D.~A.}
  \bibnamefont{{Abanin}}}, \bibinfo{journal}{Phys. Rev. Lett.}
  \textbf{\bibinfo{volume}{111}}, \bibinfo{eid}{127201} (\bibinfo{year}{2013}).

\bibitem[{\citenamefont{{Chandran} et~al.}(2014)\citenamefont{{Chandran},
  {Khemani}, {Laumann}, and {Sondhi}}}]{chandran2014}
\bibinfo{author}{\bibfnamefont{A.}~\bibnamefont{{Chandran}}},
  \bibinfo{author}{\bibfnamefont{V.}~\bibnamefont{{Khemani}}},
  \bibinfo{author}{\bibfnamefont{C.~R.} \bibnamefont{{Laumann}}},
  \bibnamefont{and} \bibinfo{author}{\bibfnamefont{S.~L.}
  \bibnamefont{{Sondhi}}}, \bibinfo{journal}{\prb}
  \textbf{\bibinfo{volume}{89}}, \bibinfo{eid}{144201} (\bibinfo{year}{2014}).

\bibitem[{\citenamefont{{Grover}}(2014)}]{grover2014}
\bibinfo{author}{\bibfnamefont{T.}~\bibnamefont{{Grover}}},
  \bibinfo{journal}{ArXiv e-prints}  (\bibinfo{year}{2014}),
  \eprint{1405.1471}.

\bibitem[{\citenamefont{{Yao} et~al.}(2014)\citenamefont{{Yao}, {Laumann},
  {Gopalakrishnan}, {Knap}, {M{\"u}ller}, {Demler}, and {Lukin}}}]{yao2014}
\bibinfo{author}{\bibfnamefont{N.~Y.} \bibnamefont{{Yao}}},
  \bibinfo{author}{\bibfnamefont{C.~R.} \bibnamefont{{Laumann}}},
  \bibinfo{author}{\bibfnamefont{S.}~\bibnamefont{{Gopalakrishnan}}},
  \bibinfo{author}{\bibfnamefont{M.}~\bibnamefont{{Knap}}},
  \bibinfo{author}{\bibfnamefont{M.}~\bibnamefont{{M{\"u}ller}}},
  \bibinfo{author}{\bibfnamefont{E.~A.} \bibnamefont{{Demler}}},
  \bibnamefont{and} \bibinfo{author}{\bibfnamefont{M.~D.}
  \bibnamefont{{Lukin}}}, \bibinfo{journal}{Phys. Rev. Lett.}
  \textbf{\bibinfo{volume}{113}}, \bibinfo{eid}{243002} (\bibinfo{year}{2014}).

\bibitem[{\citenamefont{{Vasseur} et~al.}(2015)\citenamefont{{Vasseur},
  {Parameswaran}, and {Moore}}}]{vasseur2015}
\bibinfo{author}{\bibfnamefont{R.}~\bibnamefont{{Vasseur}}},
  \bibinfo{author}{\bibfnamefont{S.~A.} \bibnamefont{{Parameswaran}}},
  \bibnamefont{and} \bibinfo{author}{\bibfnamefont{J.~E.}
  \bibnamefont{{Moore}}}, \bibinfo{journal}{\prb}
  \textbf{\bibinfo{volume}{91}}, \bibinfo{eid}{140202} (\bibinfo{year}{2015}).

\bibitem[{\citenamefont{{Ros} et~al.}(2015)\citenamefont{{Ros}, {M{\"u}ller},
  and {Scardicchio}}}]{ros2015}
\bibinfo{author}{\bibfnamefont{V.}~\bibnamefont{{Ros}}},
  \bibinfo{author}{\bibfnamefont{M.}~\bibnamefont{{M{\"u}ller}}},
  \bibnamefont{and}
  \bibinfo{author}{\bibfnamefont{A.}~\bibnamefont{{Scardicchio}}},
  \bibinfo{journal}{Nucl. Phys. B} \textbf{\bibinfo{volume}{891}},
  \bibinfo{pages}{420} (\bibinfo{year}{2015}).

\bibitem[{\citenamefont{{Oganesyan} and {Huse}}(2007)}]{oganesyan2007}
\bibinfo{author}{\bibfnamefont{V.}~\bibnamefont{{Oganesyan}}} \bibnamefont{and}
  \bibinfo{author}{\bibfnamefont{D.~A.} \bibnamefont{{Huse}}},
  \bibinfo{journal}{\prb} \textbf{\bibinfo{volume}{75}}, \bibinfo{eid}{155111}
  (\bibinfo{year}{2007}).

\bibitem[{\citenamefont{{Pal} and {Huse}}(2010)}]{pal2010}
\bibinfo{author}{\bibfnamefont{A.}~\bibnamefont{{Pal}}} \bibnamefont{and}
  \bibinfo{author}{\bibfnamefont{D.~A.} \bibnamefont{{Huse}}},
  \bibinfo{journal}{\prb} \textbf{\bibinfo{volume}{82}}, \bibinfo{eid}{174411}
  (\bibinfo{year}{2010}).

\bibitem[{\citenamefont{{{\v Z}nidari{\v c}} et~al.}(2008)\citenamefont{{{\v
  Z}nidari{\v c}}, {Prosen}, and {Prelov{\v s}ek}}}]{znidaric2008}
\bibinfo{author}{\bibfnamefont{M.}~\bibnamefont{{{\v Z}nidari{\v c}}}},
  \bibinfo{author}{\bibfnamefont{T.}~\bibnamefont{{Prosen}}}, \bibnamefont{and}
  \bibinfo{author}{\bibfnamefont{P.}~\bibnamefont{{Prelov{\v s}ek}}},
  \bibinfo{journal}{\prb} \textbf{\bibinfo{volume}{77}}, \bibinfo{eid}{064426}
  (\bibinfo{year}{2008}).

\bibitem[{\citenamefont{{Canovi} et~al.}(2011)\citenamefont{{Canovi},
  {Rossini}, {Fazio}, {Santoro}, and {Silva}}}]{canovi2011}
\bibinfo{author}{\bibfnamefont{E.}~\bibnamefont{{Canovi}}},
  \bibinfo{author}{\bibfnamefont{D.}~\bibnamefont{{Rossini}}},
  \bibinfo{author}{\bibfnamefont{R.}~\bibnamefont{{Fazio}}},
  \bibinfo{author}{\bibfnamefont{G.~E.} \bibnamefont{{Santoro}}},
  \bibnamefont{and} \bibinfo{author}{\bibfnamefont{A.}~\bibnamefont{{Silva}}},
  \bibinfo{journal}{\prb} \textbf{\bibinfo{volume}{83}}, \bibinfo{eid}{094431}
  (\bibinfo{year}{2011}).

\bibitem[{\citenamefont{{Cuevas} et~al.}(2012)\citenamefont{{Cuevas},
  {Feigel'Man}, {Ioffe}, and {Mezard}}}]{cuevas2012}
\bibinfo{author}{\bibfnamefont{E.}~\bibnamefont{{Cuevas}}},
  \bibinfo{author}{\bibfnamefont{M.}~\bibnamefont{{Feigel'Man}}},
  \bibinfo{author}{\bibfnamefont{L.}~\bibnamefont{{Ioffe}}}, \bibnamefont{and}
  \bibinfo{author}{\bibfnamefont{M.}~\bibnamefont{{Mezard}}},
  \bibinfo{journal}{Nat. Commun.} \textbf{\bibinfo{volume}{3}},
  \bibinfo{eid}{1128} (\bibinfo{year}{2012}).

\bibitem[{\citenamefont{{Kj{\"a}ll} et~al.}(2014)\citenamefont{{Kj{\"a}ll},
  {Bardarson}, and {Pollmann}}}]{kjall2014}
\bibinfo{author}{\bibfnamefont{J.~A.} \bibnamefont{{Kj{\"a}ll}}},
  \bibinfo{author}{\bibfnamefont{J.~H.} \bibnamefont{{Bardarson}}},
  \bibnamefont{and}
  \bibinfo{author}{\bibfnamefont{F.}~\bibnamefont{{Pollmann}}},
  \bibinfo{journal}{Phys. Rev. Lett.} \textbf{\bibinfo{volume}{113}},
  \bibinfo{eid}{107204} (\bibinfo{year}{2014}).

\bibitem[{\citenamefont{{De Luca} and {Scardicchio}}(2013)}]{luca2013}
\bibinfo{author}{\bibfnamefont{A.}~\bibnamefont{{De Luca}}} \bibnamefont{and}
  \bibinfo{author}{\bibfnamefont{A.}~\bibnamefont{{Scardicchio}}},
  \bibinfo{journal}{EPL (Europhysics Letters)} \textbf{\bibinfo{volume}{101}},
  \bibinfo{pages}{37003} (\bibinfo{year}{2013}).

\bibitem[{\citenamefont{{Iyer} et~al.}(2013)\citenamefont{{Iyer}, {Oganesyan},
  {Refael}, and {Huse}}}]{iyer2013}
\bibinfo{author}{\bibfnamefont{S.}~\bibnamefont{{Iyer}}},
  \bibinfo{author}{\bibfnamefont{V.}~\bibnamefont{{Oganesyan}}},
  \bibinfo{author}{\bibfnamefont{G.}~\bibnamefont{{Refael}}}, \bibnamefont{and}
  \bibinfo{author}{\bibfnamefont{D.~A.} \bibnamefont{{Huse}}},
  \bibinfo{journal}{\prb} \textbf{\bibinfo{volume}{87}}, \bibinfo{eid}{134202}
  (\bibinfo{year}{2013}).

\bibitem[{\citenamefont{{Johri} et~al.}(2014)\citenamefont{{Johri},
  {Nandkishore}, and {Bhatt}}}]{johri2014}
\bibinfo{author}{\bibfnamefont{S.}~\bibnamefont{{Johri}}},
  \bibinfo{author}{\bibfnamefont{R.}~\bibnamefont{{Nandkishore}}},
  \bibnamefont{and} \bibinfo{author}{\bibfnamefont{R.~N.}
  \bibnamefont{{Bhatt}}}, \bibinfo{journal}{ArXiv e-prints}
  (\bibinfo{year}{2014}), \eprint{1405.5515}.

\bibitem[{\citenamefont{{Bardarson} et~al.}(2012)\citenamefont{{Bardarson},
  {Pollmann}, and {Moore}}}]{bardarson2012}
\bibinfo{author}{\bibfnamefont{J.~H.} \bibnamefont{{Bardarson}}},
  \bibinfo{author}{\bibfnamefont{F.}~\bibnamefont{{Pollmann}}},
  \bibnamefont{and} \bibinfo{author}{\bibfnamefont{J.~E.}
  \bibnamefont{{Moore}}}, \bibinfo{journal}{Phys. Rev. Lett.}
  \textbf{\bibinfo{volume}{109}}, \bibinfo{eid}{017202} (\bibinfo{year}{2012}).

\bibitem[{\citenamefont{{Andraschko} et~al.}(2014)\citenamefont{{Andraschko},
  {Enss}, and {Sirker}}}]{andraschko2014}
\bibinfo{author}{\bibfnamefont{F.}~\bibnamefont{{Andraschko}}},
  \bibinfo{author}{\bibfnamefont{T.}~\bibnamefont{{Enss}}}, \bibnamefont{and}
  \bibinfo{author}{\bibfnamefont{J.}~\bibnamefont{{Sirker}}},
  \bibinfo{journal}{Phys. Rev. Lett.} \textbf{\bibinfo{volume}{113}},
  \bibinfo{eid}{217201} (\bibinfo{year}{2014}).

\bibitem[{\citenamefont{{Laumann} et~al.}(2014)\citenamefont{{Laumann}, {Pal},
  and {Scardicchio}}}]{laumann2014}
\bibinfo{author}{\bibfnamefont{C.~R.} \bibnamefont{{Laumann}}},
  \bibinfo{author}{\bibfnamefont{A.}~\bibnamefont{{Pal}}}, \bibnamefont{and}
  \bibinfo{author}{\bibfnamefont{A.}~\bibnamefont{{Scardicchio}}},
  \bibinfo{journal}{Phys. Rev. Lett.} \textbf{\bibinfo{volume}{113}},
  \bibinfo{eid}{200405} (\bibinfo{year}{2014}).

\bibitem[{\citenamefont{{Hickey} et~al.}(2014)\citenamefont{{Hickey}, {Genway},
  and {Garrahan}}}]{hickey2014}
\bibinfo{author}{\bibfnamefont{J.~M.} \bibnamefont{{Hickey}}},
  \bibinfo{author}{\bibfnamefont{S.}~\bibnamefont{{Genway}}}, \bibnamefont{and}
  \bibinfo{author}{\bibfnamefont{J.~P.} \bibnamefont{{Garrahan}}},
  \bibinfo{journal}{ArXiv e-prints}  (\bibinfo{year}{2014}),
  \eprint{1405.5780}.

\bibitem[{\citenamefont{{Nanduri} et~al.}(2014)\citenamefont{{Nanduri}, {Kim},
  and {Huse}}}]{nanduri2014}
\bibinfo{author}{\bibfnamefont{A.}~\bibnamefont{{Nanduri}}},
  \bibinfo{author}{\bibfnamefont{H.}~\bibnamefont{{Kim}}}, \bibnamefont{and}
  \bibinfo{author}{\bibfnamefont{D.~A.} \bibnamefont{{Huse}}},
  \bibinfo{journal}{\prb} \textbf{\bibinfo{volume}{90}}, \bibinfo{eid}{064201}
  (\bibinfo{year}{2014}).

\bibitem[{\citenamefont{{Bar Lev} and {Reichman}}(2014)}]{barlev2014}
\bibinfo{author}{\bibfnamefont{Y.}~\bibnamefont{{Bar Lev}}} \bibnamefont{and}
  \bibinfo{author}{\bibfnamefont{D.~R.} \bibnamefont{{Reichman}}},
  \bibinfo{journal}{\prb} \textbf{\bibinfo{volume}{89}}, \bibinfo{eid}{220201}
  (\bibinfo{year}{2014}).

\bibitem[{\citenamefont{{Bauer} and {Nayak}}(2013)}]{bauer2013}
\bibinfo{author}{\bibfnamefont{B.}~\bibnamefont{{Bauer}}} \bibnamefont{and}
  \bibinfo{author}{\bibfnamefont{C.}~\bibnamefont{{Nayak}}},
  \bibinfo{journal}{J. Stat. Mech. Theor. Exp.} \textbf{\bibinfo{volume}{9}},
  \bibinfo{eid}{09005} (\bibinfo{year}{2013}).

\bibitem[{\citenamefont{{Imbrie}}(2014)}]{imbrie2014}
\bibinfo{author}{\bibfnamefont{J.~Z.} \bibnamefont{{Imbrie}}},
  \bibinfo{journal}{ArXiv e-prints}  (\bibinfo{year}{2014}),
  \eprint{1403.7837}.

\bibitem[{\citenamefont{{Serbyn} and {Moore}}(2015)}]{serbyn2015}
\bibinfo{author}{\bibfnamefont{M.}~\bibnamefont{{Serbyn}}} \bibnamefont{and}
  \bibinfo{author}{\bibfnamefont{J.~E.} \bibnamefont{{Moore}}},
  \bibinfo{journal}{ArXiv e-prints}  (\bibinfo{year}{2015}),
  \eprint{1508.07293}.

\bibitem[{\citenamefont{{Singh} et~al.}(2015)\citenamefont{{Singh},
  {Bardarson}, and {Pollmann}}}]{singh2015}
\bibinfo{author}{\bibfnamefont{R.}~\bibnamefont{{Singh}}},
  \bibinfo{author}{\bibfnamefont{J.~H.} \bibnamefont{{Bardarson}}},
  \bibnamefont{and}
  \bibinfo{author}{\bibfnamefont{F.}~\bibnamefont{{Pollmann}}},
  \bibinfo{journal}{ArXiv e-prints}  (\bibinfo{year}{2015}),
  \eprint{1508.05045}.

\bibitem[{\citenamefont{{Bar Lev} and {Reichman}}(2015)}]{barlev2015}
\bibinfo{author}{\bibfnamefont{Y.}~\bibnamefont{{Bar Lev}}} \bibnamefont{and}
  \bibinfo{author}{\bibfnamefont{D.~R.} \bibnamefont{{Reichman}}},
  \bibinfo{journal}{ArXiv e-prints}  (\bibinfo{year}{2015}),
  \eprint{1508.05391}.

\bibitem[{\citenamefont{{Deng} et~al.}(2015)\citenamefont{{Deng}, {Pixley},
  {Li}, and {Das Sarma}}}]{deng2015}
\bibinfo{author}{\bibfnamefont{D.-L.} \bibnamefont{{Deng}}},
  \bibinfo{author}{\bibfnamefont{J.~H.} \bibnamefont{{Pixley}}},
  \bibinfo{author}{\bibfnamefont{X.}~\bibnamefont{{Li}}}, \bibnamefont{and}
  \bibinfo{author}{\bibfnamefont{S.}~\bibnamefont{{Das Sarma}}},
  \bibinfo{journal}{ArXiv e-prints}  (\bibinfo{year}{2015}),
  \eprint{1508.01270}.

\bibitem[{\citenamefont{{Chen} et~al.}(2015)\citenamefont{{Chen}, {Yu}, {Cho},
  {Clark}, and {Fradkin}}}]{chen2015}
\bibinfo{author}{\bibfnamefont{X.}~\bibnamefont{{Chen}}},
  \bibinfo{author}{\bibfnamefont{X.}~\bibnamefont{{Yu}}},
  \bibinfo{author}{\bibfnamefont{G.~Y.} \bibnamefont{{Cho}}},
  \bibinfo{author}{\bibfnamefont{B.~K.} \bibnamefont{{Clark}}},
  \bibnamefont{and}
  \bibinfo{author}{\bibfnamefont{E.}~\bibnamefont{{Fradkin}}},
  \bibinfo{journal}{ArXiv e-prints}  (\bibinfo{year}{2015}),
  \eprint{1509.03890}.

\bibitem[{\citenamefont{{Pekker} et~al.}(2014)\citenamefont{{Pekker}, {Refael},
  {Altman}, {Demler}, and {Oganesyan}}}]{pekker_hilbert2014}
\bibinfo{author}{\bibfnamefont{D.}~\bibnamefont{{Pekker}}},
  \bibinfo{author}{\bibfnamefont{G.}~\bibnamefont{{Refael}}},
  \bibinfo{author}{\bibfnamefont{E.}~\bibnamefont{{Altman}}},
  \bibinfo{author}{\bibfnamefont{E.}~\bibnamefont{{Demler}}}, \bibnamefont{and}
  \bibinfo{author}{\bibfnamefont{V.}~\bibnamefont{{Oganesyan}}},
  \bibinfo{journal}{Phys. Rev. X} \textbf{\bibinfo{volume}{4}},
  \bibinfo{eid}{011052} (\bibinfo{year}{2014}).

\bibitem[{\citenamefont{{Huang}}(2015)}]{huang2015}
\bibinfo{author}{\bibfnamefont{Y.}~\bibnamefont{{Huang}}},
  \bibinfo{journal}{ArXiv e-prints}  (\bibinfo{year}{2015}),
  \eprint{1507.01304}.

\bibitem[{\citenamefont{{You} et~al.}(2015)\citenamefont{{You}, {Qi}, and
  {Xu}}}]{you2015}
\bibinfo{author}{\bibfnamefont{Y.-Z.} \bibnamefont{{You}}},
  \bibinfo{author}{\bibfnamefont{X.-L.} \bibnamefont{{Qi}}}, \bibnamefont{and}
  \bibinfo{author}{\bibfnamefont{C.}~\bibnamefont{{Xu}}},
  \bibinfo{journal}{ArXiv e-prints}  (\bibinfo{year}{2015}),
  \eprint{1508.03635}.

\bibitem[{\citenamefont{{Luitz} et~al.}(2015)\citenamefont{{Luitz},
  {Laflorencie}, and {Alet}}}]{luitz2015}
\bibinfo{author}{\bibfnamefont{D.~J.} \bibnamefont{{Luitz}}},
  \bibinfo{author}{\bibfnamefont{N.}~\bibnamefont{{Laflorencie}}},
  \bibnamefont{and} \bibinfo{author}{\bibfnamefont{F.}~\bibnamefont{{Alet}}},
  \bibinfo{journal}{\prb} \textbf{\bibinfo{volume}{91}}, \bibinfo{eid}{081103}
  (\bibinfo{year}{2015}).

\bibitem[{\citenamefont{{Potter} et~al.}(2015)\citenamefont{{Potter},
  {Vasseur}, and {Parameswaran}}}]{potter2015trans}
\bibinfo{author}{\bibfnamefont{A.~C.} \bibnamefont{{Potter}}},
  \bibinfo{author}{\bibfnamefont{R.}~\bibnamefont{{Vasseur}}},
  \bibnamefont{and} \bibinfo{author}{\bibfnamefont{S.~A.}
  \bibnamefont{{Parameswaran}}}, \bibinfo{journal}{ArXiv e-prints}
  (\bibinfo{year}{2015}), \eprint{1501.03501}.

\bibitem[{\citenamefont{{Deutsch}}(1991)}]{deutsch1991}
\bibinfo{author}{\bibfnamefont{J.~M.} \bibnamefont{{Deutsch}}},
  \bibinfo{journal}{\pra} \textbf{\bibinfo{volume}{43}}, \bibinfo{pages}{2046}
  (\bibinfo{year}{1991}).

\bibitem[{\citenamefont{{Srednicki}}(1994)}]{srednicki1994}
\bibinfo{author}{\bibfnamefont{M.}~\bibnamefont{{Srednicki}}},
  \bibinfo{journal}{\pre} \textbf{\bibinfo{volume}{50}}, \bibinfo{pages}{888}
  (\bibinfo{year}{1994}).

\bibitem[{\citenamefont{{Hosur} and {Qi}}(2015)}]{hosur2015}
\bibinfo{author}{\bibfnamefont{P.}~\bibnamefont{{Hosur}}} \bibnamefont{and}
  \bibinfo{author}{\bibfnamefont{X.-L.} \bibnamefont{{Qi}}},
  \bibinfo{journal}{ArXiv e-prints}  (\bibinfo{year}{2015}),
  \eprint{1507.04003}.

\bibitem[{\citenamefont{{Huse} et~al.}(2013)\citenamefont{{Huse},
  {Nandkishore}, {Oganesyan}, {Pal}, and {Sondhi}}}]{huse2013}
\bibinfo{author}{\bibfnamefont{D.~A.} \bibnamefont{{Huse}}},
  \bibinfo{author}{\bibfnamefont{R.}~\bibnamefont{{Nandkishore}}},
  \bibinfo{author}{\bibfnamefont{V.}~\bibnamefont{{Oganesyan}}},
  \bibinfo{author}{\bibfnamefont{A.}~\bibnamefont{{Pal}}}, \bibnamefont{and}
  \bibinfo{author}{\bibfnamefont{S.~L.} \bibnamefont{{Sondhi}}},
  \bibinfo{journal}{\prb} \textbf{\bibinfo{volume}{88}}, \bibinfo{eid}{014206}
  (\bibinfo{year}{2013}).

\bibitem[{\citenamefont{{Bahri} et~al.}(2013)\citenamefont{{Bahri}, {Vosk},
  {Altman}, and {Vishwanath}}}]{bahri2013}
\bibinfo{author}{\bibfnamefont{Y.}~\bibnamefont{{Bahri}}},
  \bibinfo{author}{\bibfnamefont{R.}~\bibnamefont{{Vosk}}},
  \bibinfo{author}{\bibfnamefont{E.}~\bibnamefont{{Altman}}}, \bibnamefont{and}
  \bibinfo{author}{\bibfnamefont{A.}~\bibnamefont{{Vishwanath}}},
  \bibinfo{journal}{ArXiv e-prints}  (\bibinfo{year}{2013}).

\bibitem[{\citenamefont{{Vosk} and {Altman}}(2014)}]{vosk2014}
\bibinfo{author}{\bibfnamefont{R.}~\bibnamefont{{Vosk}}} \bibnamefont{and}
  \bibinfo{author}{\bibfnamefont{E.}~\bibnamefont{{Altman}}},
  \bibinfo{journal}{Phys. Rev. Lett.} \textbf{\bibinfo{volume}{112}},
  \bibinfo{eid}{217204} (\bibinfo{year}{2014}).

\bibitem[{\citenamefont{{Potter} and {Vishwanath}}(2015)}]{potter2015}
\bibinfo{author}{\bibfnamefont{A.~C.} \bibnamefont{{Potter}}} \bibnamefont{and}
  \bibinfo{author}{\bibfnamefont{A.}~\bibnamefont{{Vishwanath}}},
  \bibinfo{journal}{ArXiv e-prints}  (\bibinfo{year}{2015}),
  \eprint{1506.00592}.

\bibitem[{\citenamefont{{Yao} et~al.}(2015)\citenamefont{{Yao}, {Laumann}, and
  {Vishwanath}}}]{yao2015}
\bibinfo{author}{\bibfnamefont{N.~Y.} \bibnamefont{{Yao}}},
  \bibinfo{author}{\bibfnamefont{C.~R.} \bibnamefont{{Laumann}}},
  \bibnamefont{and}
  \bibinfo{author}{\bibfnamefont{A.}~\bibnamefont{{Vishwanath}}},
  \bibinfo{journal}{ArXiv e-prints}  (\bibinfo{year}{2015}),
  \eprint{1508.06995}.

\bibitem[{\citenamefont{{Chandran}
  et~al.}(2015{\natexlab{a}})\citenamefont{{Chandran}, {Carrasquilla}, {Kim},
  {Abanin}, and {Vidal}}}]{chandran2015}
\bibinfo{author}{\bibfnamefont{A.}~\bibnamefont{{Chandran}}},
  \bibinfo{author}{\bibfnamefont{J.}~\bibnamefont{{Carrasquilla}}},
  \bibinfo{author}{\bibfnamefont{I.~H.} \bibnamefont{{Kim}}},
  \bibinfo{author}{\bibfnamefont{D.~A.} \bibnamefont{{Abanin}}},
  \bibnamefont{and} \bibinfo{author}{\bibfnamefont{G.}~\bibnamefont{{Vidal}}},
  \bibinfo{journal}{\prb} \textbf{\bibinfo{volume}{92}}, \bibinfo{eid}{024201}
  (\bibinfo{year}{2015}{\natexlab{a}}).

\bibitem[{\citenamefont{{Ponte} et~al.}(2015)\citenamefont{{Ponte},
  {Papi{\'c}}, {Huveneers}, and {Abanin}}}]{ponte2015}
\bibinfo{author}{\bibfnamefont{P.}~\bibnamefont{{Ponte}}},
  \bibinfo{author}{\bibfnamefont{Z.}~\bibnamefont{{Papi{\'c}}}},
  \bibinfo{author}{\bibfnamefont{F.}~\bibnamefont{{Huveneers}}},
  \bibnamefont{and} \bibinfo{author}{\bibfnamefont{D.~A.}
  \bibnamefont{{Abanin}}}, \bibinfo{journal}{Phys. Rev. Lett.}
  \textbf{\bibinfo{volume}{114}}, \bibinfo{eid}{140401} (\bibinfo{year}{2015}).

\bibitem[{\citenamefont{{Grover} and {Fisher}}(2014)}]{groverf2014}
\bibinfo{author}{\bibfnamefont{T.}~\bibnamefont{{Grover}}} \bibnamefont{and}
  \bibinfo{author}{\bibfnamefont{M.~P.~A.} \bibnamefont{{Fisher}}},
  \bibinfo{journal}{J. Stat. Mech. Theor. Exp.} \textbf{\bibinfo{volume}{10}},
  \bibinfo{eid}{10010} (\bibinfo{year}{2014}).

\bibitem[{\citenamefont{{Pekker} and {Clark}}(2014{\natexlab{a}})}]{pekker2014}
\bibinfo{author}{\bibfnamefont{D.}~\bibnamefont{{Pekker}}} \bibnamefont{and}
  \bibinfo{author}{\bibfnamefont{B.~K.} \bibnamefont{{Clark}}},
  \bibinfo{journal}{ArXiv e-prints}  (\bibinfo{year}{2014}{\natexlab{a}}),
  \eprint{1410.2224}.

\bibitem[{\citenamefont{{Khemani} et~al.}(2015)\citenamefont{{Khemani},
  {Pollmann}, and {Sondhi}}}]{khemani2015}
\bibinfo{author}{\bibfnamefont{V.}~\bibnamefont{{Khemani}}},
  \bibinfo{author}{\bibfnamefont{F.}~\bibnamefont{{Pollmann}}},
  \bibnamefont{and} \bibinfo{author}{\bibfnamefont{S.~L.}
  \bibnamefont{{Sondhi}}}, \bibinfo{journal}{ArXiv e-prints}
  (\bibinfo{year}{2015}), \eprint{1509.00483}.

\bibitem[{\citenamefont{{Yu} et~al.}(2015)\citenamefont{{Yu}, {Pekker}, and
  {Clark}}}]{yu2015}
\bibinfo{author}{\bibfnamefont{X.}~\bibnamefont{{Yu}}},
  \bibinfo{author}{\bibfnamefont{D.}~\bibnamefont{{Pekker}}}, \bibnamefont{and}
  \bibinfo{author}{\bibfnamefont{B.~K.} \bibnamefont{{Clark}}},
  \bibinfo{journal}{ArXiv e-prints}  (\bibinfo{year}{2015}),
  \eprint{1509.01244}.

\bibitem[{\citenamefont{{Pekker} and
  {Clark}}(2014{\natexlab{b}})}]{pekker2014_mps}
\bibinfo{author}{\bibfnamefont{D.}~\bibnamefont{{Pekker}}} \bibnamefont{and}
  \bibinfo{author}{\bibfnamefont{B.~K.} \bibnamefont{{Clark}}},
  \bibinfo{journal}{ArXiv e-prints}  (\bibinfo{year}{2014}{\natexlab{b}}),
  \eprint{1410.2224}.

\bibitem[{\citenamefont{{Friesdorf} et~al.}(2015)\citenamefont{{Friesdorf},
  {Werner}, {Brown}, {Scholz}, and {Eisert}}}]{friesdorf2015}
\bibinfo{author}{\bibfnamefont{M.}~\bibnamefont{{Friesdorf}}},
  \bibinfo{author}{\bibfnamefont{A.~H.} \bibnamefont{{Werner}}},
  \bibinfo{author}{\bibfnamefont{W.}~\bibnamefont{{Brown}}},
  \bibinfo{author}{\bibfnamefont{V.~B.} \bibnamefont{{Scholz}}},
  \bibnamefont{and} \bibinfo{author}{\bibfnamefont{J.}~\bibnamefont{{Eisert}}},
  \bibinfo{journal}{Phys. Rev. Lett.} \textbf{\bibinfo{volume}{114}},
  \bibinfo{eid}{170505} (\bibinfo{year}{2015}).

\bibitem[{\citenamefont{{Pollmann} et~al.}(2015)\citenamefont{{Pollmann},
  {Khemani}, {Cirac}, and {Sondhi}}}]{pollmann2015}
\bibinfo{author}{\bibfnamefont{F.}~\bibnamefont{{Pollmann}}},
  \bibinfo{author}{\bibfnamefont{V.}~\bibnamefont{{Khemani}}},
  \bibinfo{author}{\bibfnamefont{J.~I.} \bibnamefont{{Cirac}}},
  \bibnamefont{and} \bibinfo{author}{\bibfnamefont{S.~L.}
  \bibnamefont{{Sondhi}}}, \bibinfo{journal}{ArXiv e-prints}
  (\bibinfo{year}{2015}), \eprint{1506.07179}.

\bibitem[{\citenamefont{{Chandran}
  et~al.}(2015{\natexlab{b}})\citenamefont{{Chandran}, {Kim}, {Vidal}, and
  {Abanin}}}]{chandran2015_construct}
\bibinfo{author}{\bibfnamefont{A.}~\bibnamefont{{Chandran}}},
  \bibinfo{author}{\bibfnamefont{I.~H.} \bibnamefont{{Kim}}},
  \bibinfo{author}{\bibfnamefont{G.}~\bibnamefont{{Vidal}}}, \bibnamefont{and}
  \bibinfo{author}{\bibfnamefont{D.~A.} \bibnamefont{{Abanin}}},
  \bibinfo{journal}{\prb} \textbf{\bibinfo{volume}{91}}, \bibinfo{eid}{085425}
  (\bibinfo{year}{2015}{\natexlab{b}}).

\bibitem[{\citenamefont{{Devakul} and {Singh}}(2015)}]{devakul2015}
\bibinfo{author}{\bibfnamefont{T.}~\bibnamefont{{Devakul}}} \bibnamefont{and}
  \bibinfo{author}{\bibfnamefont{R.~R.~P.} \bibnamefont{{Singh}}},
  \bibinfo{journal}{ArXiv e-prints}  (\bibinfo{year}{2015}),
  \eprint{1508.04813}.

\bibitem[{\citenamefont{{Vosk} et~al.}(2014)\citenamefont{{Vosk}, {Huse}, and
  {Altman}}}]{vosk_theory2014}
\bibinfo{author}{\bibfnamefont{R.}~\bibnamefont{{Vosk}}},
  \bibinfo{author}{\bibfnamefont{D.~A.} \bibnamefont{{Huse}}},
  \bibnamefont{and} \bibinfo{author}{\bibfnamefont{E.}~\bibnamefont{{Altman}}},
  \bibinfo{journal}{ArXiv e-prints}  (\bibinfo{year}{2014}),
  \eprint{1412.3117}.

\bibitem[{\citenamefont{{de Roeck} et~al.}(2015)\citenamefont{{de Roeck},
  {Huveneers}, {M{\"u}ller}, and {Schiulaz}}}]{roeck2015}
\bibinfo{author}{\bibfnamefont{W.}~\bibnamefont{{de Roeck}}},
  \bibinfo{author}{\bibfnamefont{F.}~\bibnamefont{{Huveneers}}},
  \bibinfo{author}{\bibfnamefont{M.}~\bibnamefont{{M{\"u}ller}}},
  \bibnamefont{and}
  \bibinfo{author}{\bibfnamefont{M.}~\bibnamefont{{Schiulaz}}},
  \bibinfo{journal}{ArXiv e-prints}  (\bibinfo{year}{2015}),
  \eprint{1506.01505}.

\bibitem[{\citenamefont{{Mondragon-Shem}
  et~al.}(2015)\citenamefont{{Mondragon-Shem}, {Pal}, {Hughes}, and
  {Laumann}}}]{shem2015}
\bibinfo{author}{\bibfnamefont{I.}~\bibnamefont{{Mondragon-Shem}}},
  \bibinfo{author}{\bibfnamefont{A.}~\bibnamefont{{Pal}}},
  \bibinfo{author}{\bibfnamefont{T.~L.} \bibnamefont{{Hughes}}},
  \bibnamefont{and} \bibinfo{author}{\bibfnamefont{C.~R.}
  \bibnamefont{{Laumann}}}, \bibinfo{journal}{ArXiv e-prints}
  (\bibinfo{year}{2015}), \eprint{1501.03824}.

\bibitem[{\citenamefont{{Wan} et~al.}(2005)\citenamefont{{Wan}, {Sheng},
  {Rezayi}, {Yang}, {Bhatt}, and {Haldane}}}]{wan2005}
\bibinfo{author}{\bibfnamefont{X.}~\bibnamefont{{Wan}}},
  \bibinfo{author}{\bibfnamefont{D.~N.} \bibnamefont{{Sheng}}},
  \bibinfo{author}{\bibfnamefont{E.~H.} \bibnamefont{{Rezayi}}},
  \bibinfo{author}{\bibfnamefont{K.}~\bibnamefont{{Yang}}},
  \bibinfo{author}{\bibfnamefont{R.~N.} \bibnamefont{{Bhatt}}},
  \bibnamefont{and} \bibinfo{author}{\bibfnamefont{F.~D.~M.}
  \bibnamefont{{Haldane}}}, \bibinfo{journal}{\prb}
  \textbf{\bibinfo{volume}{72}}, \bibinfo{eid}{075325} (\bibinfo{year}{2005}).

\bibitem[{\citenamefont{{Sheng} et~al.}(2003)\citenamefont{{Sheng}, {Wan},
  {Rezayi}, {Yang}, {Bhatt}, and {Haldane}}}]{sheng2003}
\bibinfo{author}{\bibfnamefont{D.~N.} \bibnamefont{{Sheng}}},
  \bibinfo{author}{\bibfnamefont{X.}~\bibnamefont{{Wan}}},
  \bibinfo{author}{\bibfnamefont{E.~H.} \bibnamefont{{Rezayi}}},
  \bibinfo{author}{\bibfnamefont{K.}~\bibnamefont{{Yang}}},
  \bibinfo{author}{\bibfnamefont{R.~N.} \bibnamefont{{Bhatt}}},
  \bibnamefont{and} \bibinfo{author}{\bibfnamefont{F.~D.}
  \bibnamefont{{Haldane}}}, \bibinfo{journal}{Phys. Rev. Lett.}
  \textbf{\bibinfo{volume}{90}}, \bibinfo{eid}{256802} (\bibinfo{year}{2003}).

\bibitem[{\citenamefont{{Sheng} et~al.}(2007)\citenamefont{{Sheng}, {Sheng},
  {Haldane}, and {Balents}}}]{sheng2007}
\bibinfo{author}{\bibfnamefont{L.}~\bibnamefont{{Sheng}}},
  \bibinfo{author}{\bibfnamefont{D.~N.} \bibnamefont{{Sheng}}},
  \bibinfo{author}{\bibfnamefont{F.~D.~M.} \bibnamefont{{Haldane}}},
  \bibnamefont{and}
  \bibinfo{author}{\bibfnamefont{L.}~\bibnamefont{{Balents}}},
  \bibinfo{journal}{Phys. Rev. Lett.} \textbf{\bibinfo{volume}{99}},
  \bibinfo{eid}{196802} (\bibinfo{year}{2007}).

\bibitem[{\citenamefont{{Atas} et~al.}(2013)\citenamefont{{Atas}, {Bogomolny},
  {Giraud}, and {Roux}}}]{atas2013}
\bibinfo{author}{\bibfnamefont{Y.~Y.} \bibnamefont{{Atas}}},
  \bibinfo{author}{\bibfnamefont{E.}~\bibnamefont{{Bogomolny}}},
  \bibinfo{author}{\bibfnamefont{O.}~\bibnamefont{{Giraud}}}, \bibnamefont{and}
  \bibinfo{author}{\bibfnamefont{G.}~\bibnamefont{{Roux}}},
  \bibinfo{journal}{Phys. Rev. Lett.} \textbf{\bibinfo{volume}{110}},
  \bibinfo{eid}{084101} (\bibinfo{year}{2013}).

\bibitem[{\citenamefont{Mehta}(1991)}]{mehta1991}
\bibinfo{author}{\bibfnamefont{M.~L.} \bibnamefont{Mehta}},
  \emph{\bibinfo{title}{Random matrices}} (\bibinfo{publisher}{Academic Press},
  \bibinfo{address}{Boston, New York, San Diego}, \bibinfo{year}{1991}), ISBN
  \bibinfo{isbn}{0-12-488051-7}.

\bibitem[{\citenamefont{{Song} et~al.}(2012)\citenamefont{{Song}, {Rachel},
  {Flindt}, {Klich}, {Laflorencie}, and {Le Hur}}}]{song2012}
\bibinfo{author}{\bibfnamefont{H.~F.} \bibnamefont{{Song}}},
  \bibinfo{author}{\bibfnamefont{S.}~\bibnamefont{{Rachel}}},
  \bibinfo{author}{\bibfnamefont{C.}~\bibnamefont{{Flindt}}},
  \bibinfo{author}{\bibfnamefont{I.}~\bibnamefont{{Klich}}},
  \bibinfo{author}{\bibfnamefont{N.}~\bibnamefont{{Laflorencie}}},
  \bibnamefont{and} \bibinfo{author}{\bibfnamefont{K.}~\bibnamefont{{Le Hur}}},
  \bibinfo{journal}{\prb} \textbf{\bibinfo{volume}{85}}, \bibinfo{eid}{035409}
  (\bibinfo{year}{2012}).

\end{thebibliography}

\end{document}